\documentclass[12pt]{article}
\usepackage{amsmath}
\usepackage{amsfonts}
\usepackage{cite}

\begin{document}
\title{A possible general approach regarding  the conformability of angular observables with mathematical rules of quantum mechanics}
\author{S. Dumitru\footnote{E-mail address: s.dumitru@unitbv.ro} \\ \textit{Department of Physics, ``Transilvania'' University} \\ \textit{B-dul Eroilor 29. R-2200 Bra\c sov, Romania}}
\date{\today}

\maketitle
\begin{abstract}
The conformability of angular observales (angular momentum and azimuthal angle) with the mathematical rules of quantum mechanics is a question which still rouses debates. It is valued negatively within the existing approaches which are restricted by two amendable presumptions. If the respective presumptions are removed one can obtain a general approach in which the mentioned question is valued positively.

\end{abstract}

\section{Introduction} 
In the last decades  the pair of  angular observables $L_z$ - $\varphi$ (angular momentum - azimuthal angle) was and still is regarded as being unconformable  to the accepted mathematical rules of quantum mechanics (QM) (see \cite{1,2,3,4,5,6,7,8,9,10,11,12,13,14,15,16,17,18,19,20,21,22,23,24,25,26,27,28,29} and references). The unconformity is identified with the fact that , in some cases of circular motions, for the respective pair the Robertson - Schrodinger uncertainty relation (RSUR) is not directly applicable. That fact roused many debates and motivated various approaches planned to elucidate in an acceptable manner the missing conformability. But so far such an  elucidation was not ratified (or admited unanimously) in the scientific literature.

A minute inspection of the things shows that in the main  all the alluded approaches have a restricted character due to the presumptions (\textbf{\textit{P}}):
\begin{itemize}
\item[$\textbf{\textit{P}}_1:$] Consideration of RSUR as a twofold reference element by: (i) proscription of its direct $L_z$ - $\varphi$ descendant, and (ii) substitution of the respective descendant with some RSUR-mimic relations. $\bullet$
\item[$\textbf{\textit{P}}_2:$] consideration only of the systems with sharp circular rotations (SCR). $\bullet$
\end{itemize}
But the mentioned presumptions are amendable because they   conflict with the following facts. (\textbf{\textit{F}}):
\begin{itemize}
\item[$\textbf{\textit{F}}_1:$] From a true mathematical perspective, the RSUR is only a secondary piece, of limited validity,  resulting  from a  generally valid element represented by a Cauchy Schwarz formula (CSF). $\bullet$
\item[$\textbf{\textit{F}}_2:$] From a natural physical viewpoint the $L_z$ - $\varphi$ pair must be considered in connection not only with SCR but also with any orbital (spatial) motions ( e.g. with the non-circular rotations (NCR), presented below in section 3). $\bullet$
\end{itemize}
The above facts suggest that for the $L_z$ - $\varphi$ problem ought to search new approaches, by removing the mentioned premises $\textbf{\textit{P}}_1$ and $\textbf{\textit{P}}_2$. As we know until now such  approaches were not promoted   in the publications from the main stream of scientific literature.   In this paper we  propose a possible general approach of the mentioned kind, able to ensure a natural conformability of the $L_z$ - $\varphi$ pair with the prime mathematical rules of QM. 

For distiguinshing our proposal  from  the alluded restricted approaches, in the next section  we present briefly  the respective approaches,  including their main   assertions and a set of unavoidable shortcomings which trouble them destructively. Then, in section 3, we disclose the existence  of two examples of NCR which are in discordance with the same approaches. 

The alluded shorcomings  and discordances reenforce  the interest  for new and differently oriented approaches of the $L_z$ - $\varphi$ problem. Such an approach, of general perspective,  is argued and detailed below in our Section 4. We end the paper in Section 5 with some associate conclusions  . 

\section{Briefly on the restricted approaches }
Certainly, for the history of the $L_z$ - $\varphi$ problem,   the first  reference element was  the Robertson Schrodinger uncertainty relation (RSUR) introduced \cite {30,31} within the mathematical  formalism of QM.  In terms of usual  notations from QM the RSUR is written as 
\begin{equation}\label{eq:1}
\Delta _\psi A \cdot \Delta _\psi B \ge \frac{1}{2}\left| {\left\langle 
{\left[ {\hat A,\hat B} \right]} \right\rangle   }_\psi \right|
\end{equation}
where $\Delta _\psi A $ and ${\left\langle (....) \right\rangle   }_\psi $ signify the standard deviation of the observable $A$ respectively the mean value of $(...)$ in the state described by the wave function $\psi$, while 
$\left[ {\hat A,\hat B} \right]$ denote the commutator of the operators $\hat A$ and $\hat B$ (for more details about the notations and validity regarding the RSUR \eqref{eq:1} see the next section).

The attempts  for application of RSUR \eqref{eq:1} to the case with $A$ = $L_z$ and  $B$ = $\varphi$, i.e. to the $L_z$ - $\varphi$ pair,  evidenced the folloving intriguing facts.

On the one hand, according to the usual procedures of QM \cite{32}, the observables $L_z$ and $\varphi$ should be described by the conjugated operators
\begin{equation}\label{eq:2}
\hat L_z  =  - i\hbar \frac{\partial }{{\partial \varphi }} ,\quad \quad 
\hat \varphi  = \varphi  \cdot 
\end{equation}
respectively by the commutation relation
\begin{equation}\label{eq:3}
\left[ {\hat L_z , \hat \varphi } \right] = - i\hbar 
\end{equation}
So for the alluded pair the RSUR \eqref{eq:1} requires for its direct descendant  the relation
\begin{equation}\label{eq:4}
\Delta_\psi  L_z \cdot \Delta_\psi \, \varphi \ge \frac{\hbar}{2}
\end{equation}
On the other hand this last relation  is explicitly inapplicable in cases of angular states regarding the systems with sharp circular rotations (SCR). The respective inapplicability is pointed out here bellow.

As examples with SCR can be  quoted : (i) a particle (bead) on a circle, (ii) a 1D (or fixedaxis) rotator and (iii) non-degenerate spatial rotations. One finds examples of systems with spatial rotations in the cases of a particle on a sphere, of 2D or 3D rotators and  of an electron in a hydrogen atom respectively. The mentioned rotations are considered as non-degenerate if all the specific (orbital) quantum numbers have well-defined (unique) values.
The alluded SRC states are described  by the following wave functions taken in  a $\varphi$ - representation
\begin{equation}\label{eq:5}
\psi _m (\varphi ) = (2\pi )^{ - \frac{1}{2}} \, e^{im\varphi} 
\end{equation}
with the stipulations $\varphi \in [0, 2 \pi )$ and $m = 0, \pm 1, \pm 2,\ldots$. The respective stipulations are required by the following facts. Firstly, in cases of SRC the angle $\varphi$ is a ordinary polar coordinate which must satistfy the corresponding mathematical rules regarding the range of definition \cite{33}. Secondly, from a physical perspective, in the same cases the wave function $\psi(\varphi)$ is enforced to have  the property $\psi(0)=\psi (2 \pi - 0) := \lim \limits_{\varphi \to 2 \pi  - 0} \psi(\varphi)$.

For the alluded SRC  one finds 
\begin{equation}\label{eq:6}
\Delta _\psi  L_z  = 0, \quad \quad \Delta _\psi \,\varphi = \frac{\pi }{{\sqrt 3 }}
\end{equation}
But these expressions for $\Delta _\psi  L_z$ and $\Delta _\psi \varphi$ are incompatible with relation \eqref{eq:4}.

For avoiding the mentioned incompatibility many publications promoted the conception that in the case  of
$L_z - \varphi$  pair the RSUR \eqref{eq:1} and the associated  procedures of QM do not work correctly. Consequently it was accredited the idea that formula \eqref{eq:4} must be proscribed and replaced by adjusted $\Delta _\psi L _z - \Delta _\psi\, \varphi$ relations planned to mime the RSUR \eqref{eq:1}. So, along the years, a lot of such  mimic relations were proposed. In the main the respective relations can be expressed in one of the following forms:
\begin{equation}\label{eq:7}
\frac{{\Delta _\psi  L_z  \cdot \Delta _\psi  \varphi }}{{a\left( {\Delta _\psi  \varphi } \right)}} \ge \hbar \left| {\left\langle {b\left( \varphi  \right)} \right\rangle _\psi  } \right|
\end{equation}
\begin{equation}\label{eq:8}
\Delta _\psi  L_z  \cdot \Delta _\psi  f(\varphi ) \ge \hbar \,
\left| {\left\langle {g(\varphi )} \right\rangle _\psi  } \right|
\end{equation}
\begin{equation}\label{eq:9}
\left( {\Delta _\psi  L_z } \right)^2  + \hbar ^2 \left( {\Delta _\psi u(\varphi )} \right)^2 \ge \hbar ^2 
\left\langle {v(\varphi )} \right\rangle _\psi ^2 
\end{equation} 
\begin{equation}\label{eq:10}
\Delta _\psi  L_z \cdot \Delta _\psi  \,\varphi \ge 
\frac{\hbar }{2} \left| {1 - 2\pi \left| {\psi (2 \pi  - 0)} \right|} \right|
\end{equation} 
In \eqref{eq:8} - \eqref{eq:9} by $a, b, f, g, u$ and $v$ are  denoted various adjusting functions ( of $\Delta _\psi  \varphi$ or of $\varphi$), introduced in literature by means of some circumstantial (and more or less fictitious) considerations. 

Among the relations \eqref{eq:7} - \eqref{eq:10} of some popularity is (8) with $f(\varphi)=sin\varphi\;(or = cos\varphi)$ respectively $g(\varphi)=[\hat L_z,f(\hat\varphi)]$.  But, generally speaking, none of the respective relations is agreed unanimously as a suitable model able to substitute formula \eqref{eq:4}.

A minute examination of the facts shows that, in essence, the relations \eqref{eq:7} -\eqref{eq:10} are troubled by shortcomings revealed within the following remarks (\textbf{\textit{R}}):
\begin{itemize}
\item[$\textbf{\textit{R}} _ 1:$] The relation \eqref{eq:10} is correct from the usual perspective of QM  (see  formulas \eqref{eq:18} and \eqref{eq:21} in the next secion). But the respective relation evidently does not mime   the RSUR \eqref{eq:1} presumed as standard within the mentioned restricted approaches of $L_z - \varphi$ problem. $\bullet$
\item[$\textbf{\textit{R}} _ 2:$] Each replica from the classes depicted by  \eqref{eq:7} -\eqref{eq:10} were planned to harmonize in a mimic fashion with the same  presumed reference element  represented by RSUR \eqref{eq:1}. But, in  spite of such plannings, regarded comparatively, the respective replicas    are not mutually equivalent. $\bullet$
\item[$\textbf{\textit{R}}_3:$] Due to the absolutely circumstantial considerations by which they are introduced, the relations  \eqref{eq:7} - \eqref{eq:9} are in fact ad hoc formulas without any direct descendence from general mathematics of QM. Consequently the respective relations ought to be appeciated by taking into account sentences such are:

\textit{"`In ...science, ad hoc often means the addition of corollary hypotheses or adjustment to a ... scientific theory to save the theory from being falsified by compensating for anomalies not anticipated by the theory in its unmodified form.  ...Scientists are often suspicious or skeptical of theories that rely on ...  ad hoc adjustments "'} \cite{34}

Then, if one wants to preserve the mathematical formalism of QM as a unitary theory, as it is accreditated in our days, the relations \eqref{eq:7} -  \eqref{eq:9} must be regarded as unconvincing and inconvenient (or even prejudicial)      elements. $\bullet$
\item[$\textbf{\textit{R}}_4:$] In fact in relations \eqref{eq:7} - \eqref{eq:9} the angle $\varphi$ is substituted more or less factitiously with the adjusting functions $a, b, f, g, v$ or $u$. Then in fact , from a natural perspective of physics, such  substitutions, and consequently the respective relations, are only  mathematical artifacts. But, in physics,  the mathematical artifacts burden the scientific discussions by additions of extraneous entities (concepts, assertions, reasonings, formulas) which are not associated with a true  information regarding the real world. Then, for a good efficiency of the discussions, the alluded additions ought to be evaluated by taking into account the principle of parsimony: \textit{"`Entities should not be multiplied unnecessarily"'}(known also \cite{35,36} as the   \textit{"`Ockham's Razor"'} slogan). Within such an evaluation the relations \eqref{eq:7} - \eqref{eq:9} appear as unnecessary exercises  which  do not give  a real and useful contribution for the elucidation of  the $L_z$ - $\varphi$ problem. $\bullet$
\end{itemize}

In  our opinion  the facts revealed in this section   offer a minimal but sufficient base for concluding that  as regards the $L_z$ - $\varphi$ problem the approaches  restricted around the premises $\textbf{\textit{P}}_1$ and 
$\textbf{\textit{P}}_2$ are unable to offer   true  and natural solutions. 
\section{The discordant examples with non-circular rotations}
The  discussions presented in the previous section regard the situation of the $L_z$ - $\varphi$ pair in relation  with the  mentioned SCR. But  here is the place to note  that the same pair must be considered also in connection with other orbital (spatial) motions which  differ from SCR.  Such motions are the non-circular rotations (NCR) . As examples of NCR we mention   the quantum torsion pendulum (QTP) respectively   the degenerate spatial rotations of the systems  mentioned  in the previous section (i.e. a particle on a sphere,  2D or 3D rotators and an electron in a hydrogen atom). A rotation (motion) is degenerate if the energy of the system is well-specified while the non-energetic quantum numbers (here of orbital nature) take all  permitted values.

From the class of NCR let us firstly refer to the case of a QTP which in fact is  a simple quantum oscillator. Indeed a QTP which oscillates around the z-axis is characterized by the Hamiltonian 
\begin{equation}\label{eq:11}
\hat H = \frac{1}{{2I}} \hat L_z^2  + \frac{1}{2} J\omega ^2 \varphi ^2 
\end{equation}
Note that in this expression  $\varphi$ denotes the azimuthal angle whose range of definition is the interval $(-\infty, \infty)$. In the same exppression  appears $\hat{L}_z$ as the $z$-component of angular momentum operator defined also by \eqref{eq:2}. The other symbols  $J$ and $\omega$ in \eqref{eq:11} represent the QTP momentum of inertia respectively the frequency  of torsional oscillations.

The Schrodinger equation  associated to the Hamiltonian \eqref{eq:11} shows that the QTP have eigenstates described 
by the wave functions
\begin{equation}\label{eq:12}
\psi _n (\varphi ) = \psi _n (\xi )\; \propto \;\exp \left( { - \frac{{\xi ^2 }}{2}} \right)
\mathcal{H}_n (\xi )\,,\quad \xi  = \varphi \sqrt{\frac{{J\omega }}{\hbar }} 
\end{equation}
where $n = 0, 1, 2, 3,\ldots$ signifies the oscillation quantum number and $\mathcal{H}_n(\xi)$ 
stand for Hermite polinomials of $\xi$. The eigenstates described by \eqref{eq:12}  have   energies $E_n = \hbar\omega(n + 1/2)$.\\ 
In the states \eqref{eq:12} for the observables $L_z$and $\varphi$ associated with the operators \eqref{eq:2} one obtains the expressions
\begin{equation}\label{eq:13}
\Delta _\psi  L_z  = \sqrt {\hbar J\omega \left( {n + \frac{1}{2}} 
\right)} \, , \quad \quad \Delta _\psi  \varphi  = \sqrt {\frac{\hbar }
{{J\omega }}\left( {n + \frac{1}{2}} \right)} 
\end{equation}
which  are completely similar with the corresponding ones for the $x - p$ pair of a rectiliniar oscillator \cite {32}.

 With the expressions \eqref{eq:13} for $\Delta _\psi L_z$ and $\Delta _\psi \varphi$ one finds that in the case of  QTP the $L_z - \varphi$ pair satisfies the proscribed formula \eqref{eq:4}.

From the same class of NCR let us now refer to a degenerate state of a particle on a sphere or of a 2D rotator. In such a state the energy is $E= \hbar^2 l (l+ 1)/2J$ where the orbital number $l$ has a well-defined value ($J$ = moment of inertia). In the same state the magnetic number $m$ can take all the values $-l, -l+1,\ldots, -1, 0, 1, \ldots,l-1, l$. Then the mentioned state is described by a wave function of the form
\begin{equation}\label{eq:14} 
\psi _l (\theta , \varphi ) = \sum\limits_{m =  - l}^l {c_m } \,Y_{lm} (\theta ,\varphi )
\end{equation}
Here $\theta$ and $\varphi$ denote  polar respectively  azimuthal angles 
( $\theta \in  [0,\pi],  \varphi \in [0, 2\pi)$), $Y_{lm} \,(\theta ,\varphi )$ are the spherical functions and $c_m $ represent complex coefficients which satisfy the normalization condition $\sum\limits_{m =  - l}^l {\left| {c_m } \right|^2 } = 1$. With the expressions \eqref{eq:2} for the operators $\hat L _z $ and $\hat \varphi $ in a state described by \eqref{eq:14} one obtains
\begin{equation}\label{eq:15}
\left( \Delta_\psi   L_z  \right)^2 = \sum\limits_{m =  - l}^l {\left| {c_m } \right|^2 } \,
\hbar^2 \,  m^2  - \left[ \sum_{m =  - l}^l \left| {c_m } \right|^2 \hbar m  \right]^2 
\end{equation}
\begin{eqnarray}\label{eq:16}
\left( \Delta_\psi  \,\varphi  \right)^2  &=& \sum\limits_{m =  - l}^l 
\sum\limits_{r =  - l}^l c_m^*  \, c_r \left( Y_{lm} , \varphi^2 \, Y_{lr}  \right) - \nonumber \\
&-& \left[ \sum\limits_{m =  - l}^l \sum\limits_{r =  - l}^l c_m^* \, c_r  \left(
Y_{lm} ,\varphi Y_{lr}  \right)  \right]^2   
\end{eqnarray}
where $(f, g)$ denotes the scalar product of the functions $f$ and $g$.

By means of the expressions \eqref{eq:15} and \eqref{eq:16} one finds that in the case of alluded NCR described by the wave functions \eqref{eq:14} it is possible for the proscribed formula \eqref{eq:4} to be satisfied. Such a possibility is conditioned by the concrete values of the   coefficients $c _m$.

Now is the place for the following remark
\begin{itemize}
\item[$\textbf{\textit{R}} _ 5:$] As regards the $L_z$ - $\varphi$ problem, due to  the here revealed aspects,  the NCR examples  exceed the bounds of the presumptions $\textbf{\textit{P}}_1$ and $\textbf{\textit{P}}_2$ of usual restricted approaches. That is why the mentioned problem  requires new  approaches of general nature if it is possible. $\bullet$
\end{itemize}

\section{A possible general appoach and some remarks associated with it.}
A general approach of the $L_z$ - $\varphi$ problem, able to avoid the shortcomings and discordances revealed in the previous two sections, must be done by starting from the prime  mathematical rules of QM. Such an approach is possible to be obtained as follows.

Let us  appeal to the usual concepts and notations of QM. We consider a quantum system whose state (of orbital nature) 
 and two observables $A _j$ ($j = 1, 2$) are described by the wave function $\psi $ respectively by the operators $\hat A_j  $.  As usual with  $(f,g)$ we denote the scalar product of the functions $f$ and $g$ . In relation with the mentioned state,    the quantities $\left\langle A_j  \right\rangle _\psi = \left( {\psi \,, \hat A_j \psi } \right)$ and $\delta _\psi  \hat A_j  = \hat A_j  - \left\langle {\hat A_j } \right\rangle _\psi  $ represent the mean (expected) value respectively the deviation-operator of the observable $A_j$ regarded as a random variable. Then, by taking 
 $A_1 = A$ and $A_2 = B$, for the two observables can be written the following Cauchy-Schwarz relation:
\begin{equation}\label{eq:17}
\left( {\delta _\psi  \hat A \psi ,\,\delta _\psi  \hat A\psi } \right)\left( 
{\delta _\psi  \hat B\psi ,\,\delta _\psi  \hat B\psi } \right) \ge \left| 
{\left( {\delta _\psi  \hat A\psi ,\,\delta _\psi  B\psi } \right)} \right|^2 
\end{equation}
For an observable $A_j$ regarded as a random variable the quantity $ \Delta _\psi A_j = \left( {\delta _\psi  \hat A_j \psi ,\,\delta _\psi  \hat A_j \psi } \right)^{\frac{1}{2}}$ represents its standard deviation. From \eqref{eq:17} it results directly that the standard deviations $\Delta _\psi A$ and  $\Delta _\psi B$ of the  observables $A$ and $B$ satisfy the relation
\begin{equation}\label{eq:18}
\Delta _\psi  A \cdot \Delta _\psi  B \ge \left| {\left( {\delta _\psi  
\hat A \psi ,\,\delta _\psi  B \psi } \right)} \right|
\end{equation}
which can be called \emph{Cauchy-Schwarz formula} (CSF).

Note that CSF \eqref{eq:18} (as well as the relation \eqref{eq:17}) is always valid,   i.e. for all observables, systems and states. Add here the important observation that the CSF \eqref{eq:18} implies the restricted  RSUR \eqref{eq:1} only in the cases when the two operators $\hat A  = \hat A_1$ and $\hat B = \hat A_2$ satisfy the conditions
\begin{equation}\label{eq:19}
\left( {\hat A_j \psi ,\hat A_k \psi } \right) = \left( {\psi ,\hat A_j 
\hat A_k \psi } \right)\quad \quad \left( {j = 1,2;\;k = 1,2} \right)
\end{equation}
Indeed in such cases one can write the relation
\begin{eqnarray}\label{eq:20}
\left( {\delta _\psi  \hat A\psi , \, \delta _\psi  \hat B\psi } \right) 
&=& \frac{1}{2} \left( {\psi , \, \left( {\delta _\psi  \hat A \cdot \delta _\psi 
\hat B\psi  + \delta _\psi  \hat B \cdot \delta _\psi  \hat A} \right)\psi } \right) - \nonumber \\
&-& \frac{i}{2}\left( {\psi , \, i\left[{\hat A, \, \hat B} \right]\psi} \right) 
\end{eqnarray}
where the two terms from the right hand side are purely real and  imaginary  quantities respectively. Therefore in the mentioned cases from \eqref{eq:18} one finds
\begin{equation}\label{eq:21}
\Delta _\psi  A \cdot \Delta _\psi  B \ge \frac{1}{2}\left| 
{\left\langle {\left[ {\hat A,\, \hat B} \right]} \right\rangle _\psi  } \right|
\end{equation}
i.e. the well known RSUR \eqref{eq:1}.

The above general framing of RSUR \eqref{eq:1}/\eqref{eq:21} shows that for the here investigated question of
$L_z$ - $\varphi$ pair it is important to examine the fulfilment of the conditions \eqref{eq:19} in each of the considered case. In this sense the following remarks are of direct interest.
\begin{itemize}
\item[$\textbf{\textit{R}} _ 6:$] In the cases described by the wave functions \eqref{eq:5} for $L _z -\varphi$ pair one finds 
\begin{equation}\label{eq:22}
\left( {\hat L_z \psi _m ,\,\hat \varphi \psi _m } \right) = \left( 
{\psi _m ,\,\hat L_z \hat \varphi \psi _m } \right) + i\hbar 
\end{equation}
i.e. a clear violation in respect with the conditions \eqref{eq:19} $\bullet$ 
\item[$\textbf{\textit{R}} _ 7:$] In the cases associated with the wave functions \eqref{eq:12} and \eqref{eq:14} 
for $L _z - \varphi$ pair one obtains
\begin{equation}\label{eq:23}
\left( {\hat L_z \psi _n ,\,\hat \varphi \psi _n } \right)= \left( {\psi _n ,\,\hat L_z 
\hat \varphi \psi _n } \right)
\end{equation}
\begin{eqnarray}\label{eq:24}
&&\left( {\hat L_z \psi _l ,\,\hat \varphi \psi _l } \right) = \left(
{\psi _l ,\,\hat L_z \hat \varphi \psi _l } \right) +  \nonumber \\
&&+ i\hbar \left\{ 1+ 2\, \mathrm{Im} \left[ 
{\sum\limits_{m =  - l}^l {\sum\limits_{r =  - l}^l {c_m^* \, c_r \,\hbar m\left(
{Y_{lm} ,\,\hat \varphi \,Y_{lr} }\right)} } } \right] \right\} 
\end{eqnarray}
(where Im $[\alpha]$ denotes the imaginary part of $\alpha$). $\bullet$
\item[$\textbf{\textit{R}} _ 8:$] For any wave function $\psi (\varphi)$ with 
$\varphi \in [0, 2 \pi )$ and $\psi (2 \pi - 0) = \psi (0)$ it is generally true the formula
\begin{equation}\label{eq:25}
\left| {\left( {\delta _\psi  \hat L_z \,\psi ,\,\delta _\psi  \hat \varphi \,\psi } 
\right)} \right| \ge \frac{\hbar }{2} \left| {1 - 2 \pi \left| {\psi 
\left( {2 \pi  - 0} \right)} \right|} \right|
\end{equation}
which together with CSF \eqref{eq:18} confirms  relation \eqref{eq:10}. $\bullet$
\end{itemize}
The things mentioned above in this section justify the following remarks
\begin{itemize}
\item[$\textbf{\textit{R}} _ 9:$] The CSF \eqref{eq:18} is an ab origine element in respect with the  RSUR \eqref{eq:1}/\eqref{eq:23}. Moreover, \eqref{eq:18} is always valid, independently if the conditions \eqref{eq:19} are fulfilled or not. $\bullet$
\item[$\textbf{\textit{R}} _ {10}:$] The usual RSUR \eqref{eq:1}/\eqref{eq:21} are valid only in the circumstances strictly delimited by the conditions \eqref{eq:19} and they are false in all other situations. $\bullet$
\item[$\textbf{\textit{R}} _ {11}:$] Due to the relations \eqref{eq:22}  in the cases described by the wave functions \eqref{eq:5}  the conditions \eqref{eq:19} are not fulfilled. Consequently in such cases the restricted RSUR 
\eqref{eq:1}/\eqref{eq:21} are essentially inapplicable for the pairs $L _z - \varphi$ . However one can see that in the respective cases, mathematically,  the CSF \eqref{eq:18} remains valid as a trivial equality $0 = 0$. $\bullet$
\item[$\textbf{\textit{R}} _ {12}:$] In the cases of NCR described by \eqref{eq:12} the $L _z - \varphi$ pair satisfies the conditions \eqref{eq:19} (mainly due to the relation \eqref{eq:23}). Therefore in the respective cases the  RSUR \eqref{eq:1}/\eqref{eq:21} are valid for $L _z$ and $\varphi$. $\bullet$
\item[$\textbf{\textit{R}} _ {13}:$] The fulfilment of the conditions \eqref{eq:19}  by the $L _z -\varphi$ pair for the NCR associated with \eqref{eq:14} depends on the annulment of the second term in the right hand side  from \eqref{eq:24} (i.e. on the values of the coefficients $c _m$). Adequately, in such a case,  the correctness of the corresponding RSUR \eqref{eq:1}/\eqref{eq:21} shows the same dependence. $\bullet$
\item[$\textbf{\textit{R}} _ {14}:$] The result \eqref{eq:25} points out the fact that the adjusted relation \eqref{eq:10} 
 is only a secondary piece derivable fom the generally valid CSF \eqref{eq:18}. $\bullet$
\item[$\textbf{\textit{R}} _ {15}:$] The mimic relations \eqref{eq:7} - \eqref{eq:9}  regard the cases with SCR described by the wave functions \eqref{eq:5} when $\varphi$ plays the role of polar coordinate. But for such a role \cite{32} in order to be a unique (univocal) variable $\varphi$ must be defined naturally only inthe range $[0, 2 \pi)$. The same range is considered in practice for the normalization of the wave functions \eqref{eq:5}. Therefore, in the cases under discussion the derivative with respect to $\varphi$ refers to the mentioned  range. Particularly for the extremities of the interval $[0, 2 \pi)$ it has to operate with backward respectively forward derivatives. So in the alluded SCR cases the relations \eqref{eq:3} and \eqref{eq:4} act well, with a natural correctness. The same correctness is shown by the respective relations in connection with the NCR described by the wave functions \eqref{eq:12} or \eqref{eq:14}.
In fact, from a more general perspective, the relations \eqref{eq:2} and \eqref{eq:3} regard the QM operators $\hat L _z$ and $\hat \varphi$. Therefore they must have unique forms - i.e. expressions which do not depend on the particularities of the considered situations (e.g. systems with SCR or with NCR). $\bullet$
\item[$\textbf{\textit{R}} _ {16}:$] The troubles of RSUR \eqref{eq:1} regarding $L _z - \varphi$ pair are directly connected with the conditions \eqref{eq:19}. Then it is strange that in almost all the QM literature the respective conditions are not taken into account adequately. The reason seems to be related with  the nowadays dominant Dirac's $<bra|$ and $|ket>$ notations. In the respective notations  the terms from the both sides of \eqref{eq:19} have a unique representation namely $ < \psi |\hat A_j \,\hat A_k |\psi  > $. The respective uniqueness can entail confusion (unjustified supposition) that the  conditions \eqref{eq:19} are always fulfiled. It is interesting to note that systematic investigations on the confusions/surprises generated by the Dirac's notations were started only recently \cite{37}. Probably that further efforts on the line of such investigations will bring a new light on the conditions \eqref{eq:19} as well as on other QM questions. $\bullet$
\end{itemize}
The ensemble of things presented above in this section  appoints a possible general approach for the discussed $L_z$ - $\varphi$ problem and answer to a number of questions associated with the respective problem. Some significant aspects of the respective approach are noted in the next section.

\section{Conclusions}
The facts and arguments discussed in the previous sections guide to the following conclusions $(\textbf{\textit{C}})$:
\begin{itemize}
\item[$\textbf{\textit{C}} _ 1:$] For the $L _z - \varphi$ pair the relations \eqref{eq:2} - \eqref{eq:3} are always viable in respect with the general CSF \eqref{eq:18}. That is why, from the QM perspective,  for a correct description of questions regarding the respective pair, it is not at all necessary to resort to the mimetic formulas \eqref{eq:7} - \eqref{eq:10}. Eventually the respective formulas  can be accounted as ingenious execises of pure mathematical facture. An adequate description of the mentioned kind can be given by taking CSF \eqref{eq:18} and associated QM procedures  as basic elements. $\bullet$ 
\item[$\textbf{\textit{C}} _ 2:$] In respect with the conjugated observables $L _z$ and $ \varphi$  the RSUR \eqref{eq:1}/\eqref{eq:21} is not adequate for the role of reference element for normality . For such a role the CSF \eqref{eq:18} is the most suitable. In some cases of interest  the respective CSF degenerates in the trivial equality $0 = 0$. $\bullet$ 
\item[$\textbf{\textit{C}} _ 3:$] In reality the usual procedures of QM ( illustrated above by the relations \eqref{eq:2}, \eqref{eq:3}, \eqref{eq:17} and \eqref{eq:18}) work well and without anomalies in all situations regarding the 
$L_z - \varphi$ pair. Consequently  with regard to the conceptual as well as practical interests of science the mimic relations like \eqref{eq:7} - \eqref{eq:9} appear as useless inventions. $\bullet$
\end{itemize}

Now we wish to add the following observations $(\textbf{\textit{O}})$:
\begin{itemize}
\item[$\textbf{\textit{O}}_1:$] Mathematically the relation \eqref{eq:17} is generalisable in the form
\begin{equation}\label{eq:26}
 \det \left[ {\left( {\delta _\psi  \hat A_j 
\psi ,\,\delta _\psi  \hat A_k \psi } \right)} \right] \ge 0
\end{equation}
where $\det \left[ {\alpha_{jk} } \right]$  denotes the determinant with elements  $\alpha _{jk}$ and $j=1,2,...,r ;\: k=1,2,...,r $ with $r\geq 2$. Such a form results from the fact that  the quantities $\left( {\delta _\psi  \hat A_j \psi ,\,\delta _\psi  \hat A_k \psi } \right)$   constitute the elements of  a Hermitian and non-negatively defined matrix. Newertheless, comparatively with \eqref{eq:17},  the generalisation  \eqref{eq:26} does not bring supplementary and inedited features regarding the conformability of observables $L_z$ - $\varphi$ with the mathematical rules of QM. $\bullet$
\item[$\textbf{\textit{O}}_2:$] We consider that the above considerations about the problem  of $L_z$ - $\varphi$ pair can be of some non-trivial interest for a possible revised approach of the similar problem of the pair $N$ - $\phi$ (number - phase) which is  debated in  a somewhat similar manner in recent publications ( see \cite{9,16,17,18,38,39,40,41,42,43} and references). $\bullet$
\item[$\textbf{\textit{O}}_3:$] Note that we have limited this paper only to mathematical aspects associated with the RSUR \eqref{eq:1} , without incursions in debates about the interpretations of the respective RSUR. Some opinions about the respective interpretations and connected questions are given in \cite{44}. But the subject is delicate and probably that it will rouse further debates. $\bullet$
\end{itemize}

\end{document}